\documentclass[prd, 10pt,aps,nofootinbib, floatfix, notitlepage,twocolumn,longbibliography]{revtex4-1}
              
\usepackage{amsmath,amsfonts,amssymb}
\usepackage{mathrsfs}
\usepackage{graphicx}
\usepackage[english]{babel} 
\usepackage{ulem}
\usepackage{url} 
\usepackage{color}
\usepackage{bbold}
\usepackage{adjustbox}
 \usepackage[utf8]{inputenc} 
\usepackage[colorlinks=true,citecolor=blue,urlcolor=blue]{hyperref}

\newcommand{\nef}{\Delta N_\text{eff}}
\newcommand{\beq}{\begin{equation}}
\newcommand{\eeq}{\end{equation}}
\newcommand{\bea}{\begin{eqnarray}}
\newcommand{\eea}{\end{eqnarray}}

\renewcommand{\(}{\left(}
\renewcommand{\)}{\right)}

\begin{document}

\title{The DFSZ axion in the CMB}

\author{Ricardo Z.~Ferreira}
\email{rzambujal@ifae.es}
\affiliation{Nordita, KTH Royal Institute of Technology and Stockholm University, Roslagstullsbacken 23, SE-106 91 Stockholm, Sweden \\ and \\ Institut de F\'isica d'Altes Energies (IFAE) and 	The Barcelona Institute of Science and Technology (BIST),	Campus UAB, 08193 Bellaterra, Barcelona }
\author{Alessio Notari}
\email{notari@fqa.ub.edu}
\affiliation{Departament de F\'isica Qu\`antica i Astrofis\'ica \& Institut de Ci\`encies del Cosmos (ICCUB), Universitat de Barcelona, Mart\'i i Franqu\`es 1, 08028 Barcelona, Spain}
\author{Fabrizio Rompineve}
\email{fabrizio.rompineve@tufts.edu}
\affiliation{Institute of Cosmology, Department of Physics and Astronomy, Tufts University, Medford, MA 02155, USA
\looseness=-1}

\date{\today}

\begin{abstract}

\noindent We perform for the first time a dedicated analysis of cosmological constraints on DFSZ QCD axion models. Such constructions are  especially interesting in light of the recent Xenon-1T excess and of hints from stellar cooling. 
In DFSZ models, for $m_a\gtrsim 0.1$ eV, scatterings of pions and muons can produce a sizable cosmic background of thermal axions, that behave similarly to massive neutrinos. However, the pion coupling  
depends on the alignment between the vevs of two Higgs doublets, and can be significantly suppressed or enhanced with respect to the KSVZ scenario. Using the latest Planck and BAO data, we find $m_a\leq 0.2~\text{eV}$ at $95\%$ C.L., when the axion coupling to pions $c_{a\pi}$ is maximal. Constraints on $m_a$, instead, can be significantly relaxed when $c_{a\pi}$ is small. In particular, we point out that in the so-called DFSZ-II model, where the axion coupling to leptons does not vanish simultaneously with $c_{a\pi}$, production via muons gives $m_a\leq 0.6~\text{eV}$ at $95\%$ C.L., whereas in the DFSZ-I model bounds on $m_a$ can be fully lifted.
We then combine cosmological data with recent hints of a DFSZ axion coupled to electrons from the Xenon-1T experiment, finding in this case that the axion mass is constrained to be in the window $0.07 ~\text{eV} \lesssim m_a \lesssim 1.8\, (0.3)~\text{eV}$ for the DFSZ-I (DFSZ-II) model. A similar analysis with stellar cooling hints gives $3  ~\text{meV} \lesssim m_a \lesssim 0.2 ~\text{eV}$ for DFSZ-II, while no constraint arises in the DFSZ-I case. Forthcoming CMB Stage 4 experiments will be able to further test such scenarios; for instance the Xenon-1T window should be fully probed at $2\sigma$ for a DFSZ-I axion.

\end{abstract}

\maketitle

\section{Introduction}

The QCD axion~\cite{Weinberg:1977ma, Wilczek:1977pj} is arguably among the best motivated hypothetical addition to the Standard Model (SM) of particles physics. By coupling to QCD, it provides an elegant solution to the CP problem of strong interactions, while its oscillations could explain the observed dark matter in the Universe. However, its experimental detection is inevitably very challenging, due to its very weak interactions with the SM. It is thus of crucial importance to investigate complementary probes of the QCD axion, such as cosmological and astrophysical observations, and to combine them with current laboratory searches.

The possibility to detect imprints of the QCD axion in the Cosmic Microwave Background (CMB) is particularly exciting. Indeed, through the weak but unavoidable interactions with the SM, QCD axion quanta can be thermally produced above~\cite{Turner:1986tb,Masso:2002np,Graf:2010tv,Salvio:2013iaa}, across \cite{Arias-Aragon:2020shv} and/or below~\cite{Brust:2013xpv,Baumann:2016wac,Ferreira:2018vjj, DEramo:2018vss} the ElectroWeak Phase Transition (EW PT). Depending on the axion mass, this can lead to a population of axions that remain relativistic up to the epoch of recombination, in contrast to the non-relativistic population generated by the misalignment mechanism and by the decay of topological defects~(see~\cite{Gorghetto:2018myk, Hindmarsh:2019csc, Gorghetto:2020qws} for recent numerical estimates of this latter contribution). Such a thermally produced population would then act as dark radiation and eventually as warm dark matter. The CMB is sensitive to the fraction of energy density in these components, therefore constraints on the QCD axion mass and couplings can be in principle derived with CMB observations. In particular, the amount of dark radiation at recombination is commonly expressed in terms of the effective number of extra neutrino species beyond the three SM ones, $\Delta N_\text{eff}$, with current CMB and BAO data imposing $\Delta N_\text{eff}<0.296$~\cite{Aghanim:2018eyx} at $95\%$ C.L.~(the addition of Pantheon and SH$_0$ES supernovae measurements favors a non-vanishing $\Delta N_\text{eff}$ at slightly more than $2\sigma$ level~\cite{Ballesteros:2020sik, Gonzalez:2020fdy} $\Delta N_\text{eff}=0.34^{+0.16}_{-0.15}$). The abundance of a relic species that decouples from the early Universe plasma at a temperature $T_D$ is suppressed by the total number of degrees of freedom of the plasma at decoupling, and thus it becomes larger if  $T_D$ decreases. Therefore, the largest relic axion abundance arises when thermalization occurs below the QCD phase transition (QCD PT), through scattering of the lightest particles: leptons~\cite{Turner:1986tb,DEramo:2018vss} and pions~\cite{Chang:1993gm,Hannestad:2005df}. For this to be the case, the axion mass $m_a$ should be larger than about $\sim 0.05$~eV and correspondingly the axion decay constant $f$, that is the scale suppressing the axion couplings to SM fields, should be smaller than about $10^8$~GeV. By using CMB data and focusing on pion scattering, strong constraints on the axion mass have already been derived in one class of axion models, the KSVZ scenario~\cite{Kim:1979if,Shifman:1979if}, where the QCD axion coupling to SM fermions is negligible, since it vanishes at tree level~(see~\cite{Chang:1993gm, Hannestad:2005df} for the original works, and~\cite{Archidiacono:2013cha, DiValentino:2015wba, Millea:2020xxp,Giare:2020vzo} for recent updates using Planck 2013, 2015 and 2018 data, respectively). 

However, in an important alternative class of axion models, the DFSZ scenario~\cite{Dine:1981rt,Zhitnitsky:1980tq}, the QCD axion couples to SM fermions at tree level, thanks to the presence of an extra Higgs doublet. This opens up new possibilities: axion production from pion scatterings can be either enhanced or suppressed (see also~\cite{DiLuzio:2017ogq,Alves:2017avw}) compared to the KSVZ scenario, and production via leptons can also become important, both features depending on the details of the UV construction. The first aim of this paper is thus to pin down the range of DFSZ axion masses and couplings which are compatible with current cosmological datasets and to provide predictions for next generation CMB experiments~\cite{Abazajian:2016yjj}. To our knowledge, such an analysis has not been consistently performed to date.\footnote{Recent works~\cite{Millea:2020xxp, Giare:2020vzo} fixed the axion-pion coupling according to the KSVZ scenario (and did not discuss production via leptons). Therefore, such bounds on the QCD axion mass cannot be applied to DFSZ scenarios (especially when correlating with the Xenon-1T excess).}

Our interest in cosmological probes of the DFSZ axion is further motivated by some recent laboratory and astrophysical observations, which may be interpreted as hints of the existence of a QCD axion with a significant coupling to electrons (although these observations cannot be simultaneously addressed).  First, on the experimental front, the Xenon-1T (X1T) collaboration has recently reported an excess of electron recoil events~\cite{Aprile:2020tmw}. This can indeed be explained (with a $3.4\sigma$ significance) by solar axions interacting with electrons with a coupling $g_{ae}$, if $f/c_e \equiv m_a/g_{ae} \simeq 2\times 10^8\, \text{GeV}$ or equivalently $m_a \simeq 0.03/c_e~\text{eV}$. Since in the KSVZ model $c_e$ is loop-suppressed, the range of axion masses that can explain the X1T excess has been excluded by cosmological data long ago~\cite{Hannestad:2005df}. The situation is very different in the DFSZ scenario, where $c_e\lesssim 1/3$ leads to a suggested axion mass which sits right in the range that can be currently tested with CMB data. Second, on the astrophysical side, measurements of the cooling rates of certain white dwarfs~(see e.g.~\cite{Isern:2018uce, Corsico:2019nmr}) and red giant stars (see~\cite{Viaux:2013lha, Straniero:2018fbv, Straniero:2020iyi}) appear to hint (at slightly more than $3\sigma$) at a DFSZ axion with $m_a\sim 2/c_e~\text{meV}$, which can be probed by CMB data when $c_e\ll 1$. 

Of course the latter hint strongly disfavors the solar axion interpretation of the X1T excess~\cite{DiLuzio:2020jjp}, and the mass range $m_a > 0.02~\text{eV}$ is also known to be constrained by the neutrino signal from SN 1987A~(see~\cite{Turner:1987by, Raffelt:1987yt, Burrows:1988ah} for the original works, \cite{Chang:2018rso, Carenza:2019pxu} for the latest updates and~\cite{Bar:2019ifz} for a critical take). A weaker but possibly more robust constraint comes from horizontal branch (HB) stars, whose cooling rates generically impose $m_a\lesssim 0.3~\text{eV}$~\cite{Ayala:2014pea}.
Nonetheless, we think that there are at least two reasons to investigate the hints above with CMB data. First, cosmological and/or collider constraints are often simpler to extract and subject to more controllable uncertainties than astrophysical bounds, and thus represent important independent analyses. Second, concerning the X1T excess, more complicated axion models may exist, where constraints from star cooling can be evaded due to environmental effects (see e.g.~\cite{Bloch:2020uzh}). Third, bounds from SN 1987A can be relaxed in certain variations of the DFSZ model, where the coupling to nucleons~\cite{DiLuzio:2017ogq} and/or electrons~\cite{Bjorkeroth:2019jtx} can be suppressed. In this case, our strategy can then be used to constrain scenarios of this kind, since thermal axion production can be driven by several scattering processes, unless they are all simultaneously suppressed. 
In light of the discussion above, the second aim of this paper is thus to perform an analysis of the DFSZ axion by combining CMB data with X1T and stellar cooling hints, separately.

This paper is organized as follows: 
In Section \ref{sec:api} we briefly review the origin and properties of the axion-pion coupling.
In Section \ref{sec:deltaneff} we present and solve the Boltzmann equation for the axion abundance and provide the predictions for $\nef$. 
In Section \ref{sec:data} we discuss the details of the data analysis and present the MCMC results. 
Finally, in Section \ref{sec:conclusions} we conclude by presenting the main conclusions of our work for DFSZ axions. Additional details on scattering rates and the Boltzmann equation are provided in the Appendix.

\section{Axion interactions}

The main focus of this paper is the QCD axion, $a$, whose general Lagrangian above the scale of QCD confinement and below the EW scale is: 
\begin{eqnarray}
	\label{eq:axilag}
	\mathcal{L}_a =& & \frac{1}{2}(\partial_\mu a)^2 + \frac{\partial_\mu a}{2f} \sum_\psi  c_\psi^0 j_\psi^\mu  + \frac{\alpha_s}{8\pi}\frac{a}{f}G\tilde{G} + \mathcal{L}_{a\gamma} \nonumber \\ && - \bar{q}_L M_q q_R + \text{h.c.},
\end{eqnarray}
where $j_\psi^\mu= \bar{\psi}  \gamma^\mu \gamma^5 \bar{\psi} $ is the axial current for a generic Standard Model (SM) fermion $\psi$ (quark or lepton), $\mathcal{L}_{a\gamma} =  g_{a \gamma \gamma}^0 a F\tilde{F} /4$ is the axion coupling to photons and $M_q$ is the diagonal quark mass matrix. 
The continuous shift symmetry of the axion field is only broken by the term proportional to $G\tilde{G}$  in eq.~\eqref{eq:axilag}. Alternatively, to remove the axion coupling to gluons we may perform a chiral rotation of the quark fields, $R_a=e^{i\frac{a}{2 f}Q_a}$ with $\text{tr} \,Q_a=1$. It is convenient to define $Q_a= M_q^{-1}/\text{tr}(M_q^{-1})$, to avoid tree-level mixing with pions \cite{Georgi:1986df}. This rotation will affect the second and fourth term in eq.~\eqref{eq:axilag} and induce an axion dependence in the quark mass term:
\begin{eqnarray}
	\label{eq:axilagch}
	\mathcal{L}_a = && \frac{1}{2}(\partial_\mu a)^2 + \frac{\partial_\mu a}{2f} \sum_\psi  c_\psi j_\psi^\mu   + \mathcal{L}_{a\gamma}  - \bar{q}_L M q_R + \text{h.c.} \, , \quad
\end{eqnarray} 
where $M= R_a M_q  R_a$. The coefficients in \eqref{eq:axilagch} are given by $c_q=c_q^0-Q_a$ and the axion-photon Lagrangian is the same as above but with $g_{a\gamma\gamma}^0$ replaced by $g_{a\gamma\gamma}=g_{a\gamma\gamma}^0- 2N_c\alpha/(2\pi f)\text{Tr}(Q_aQ^2)$, where $N_c$ is the number of colors, $\alpha$ the fine-structure constant and $Q$ is the electric charge matrix of the quarks. 

As it stands, the Lagrangian \eqref{eq:axilag} requires a UV completion around the scale $f$. This is usually achieved by introducing a complex scalar field $\Phi$, with a $U(1)$ Peccei-Quinn (PQ) symmetry broken at the scale $f$ \cite{Peccei:1977hh,Peccei:1977ur}. The axion then arises as the phase of this field and the coupling to SM gluons can be generated according to two main constructions. In the KSVZ model~\cite{Kim:1979if,Shifman:1979if}, $\Phi$ is only coupled at tree-level to heavy dark fermions in the UV, therefore $c_\psi^0=0$.  In the DFSZ model~\cite{Dine:1981rt,Zhitnitsky:1980tq}, $\Phi$ couples to an extended Higgs sector with two doublets $H_u$ and $H_d$. Thus a coupling of the axion to SM quarks is automatically present and one has:
\begin{align} \label{eq:DFSZ, cu and cd}
	\textbf{KSVZ} &: \quad c^0_u= c^0_d=0 \, ,\\
	\textbf{DFSZ} &:  \quad c^0_u=\frac{1}{3}\cos^2(\beta) \, , \quad c^0_d=\frac{1}{3}\sin^2(\beta) \, ,
\end{align}
where the notation $u$ and $d$ stands for a  universal coupling to up-type and down-type quarks of any generation and $\tan\beta \equiv v_u/v_d $, where $v_{u,d}$ are the VEVs of $H_u$ and $H_d$.

Below the scale of QCD confinement, strong interactions generate a periodic potential $V(a)$ for the axion, which thereby acquires a mass:
\begin{equation}
	\label{eq:axmass}
	m_a=\frac{\sqrt{z}}{1+z}\frac{f_\pi m_\pi}{f}\simeq 0.57 \left(\frac{10^7 \text{GeV}}{f}\right)~\text{eV}.
\end{equation}
Here $z=m_u/m_d\simeq 0.47^{+0.06}_{-0.07}, f_\pi\simeq 92.3$~MeV and $m_\pi \simeq 134.98$~MeV~\cite{Zyla:2020zbs}.\footnote{Throughout this work we neglect uncertainties on the values of $f_{\pi}$ and $m_\pi$, since they do not affect our results. We use the uncertainties on $z$ as given in~\cite{Zyla:2020zbs}. Smaller uncertainties have been recently obtained from lattice simulations: $z=0.472(11)$~\cite{Gorghetto:2018ocs} and we discuss their implications for axion coupling to pions below.} 

Because of its interactions, the QCD axion field can be thermally produced in the early Universe. This leads to a population of relic axions which, depending on the axion mass, acts as a dark radiation (DR) or a warm dark matter component. Assuming relic axions decouple, their present energy density in the relativistic limit would be $\rho_{\text{DR}}=\pi^2/30 g_{a} T_a^4$, and their abundance is usually parameterized in terms of the effective number of neutrino ($\nu$) species, as (see e.g.~\cite{Baumann:2018muz})
\begin{eqnarray}
\label{eq:deltaneff}
\nonumber \Delta N_{\text{eff}}&\equiv& \dfrac{\rho_{\text{DR}}}{\rho_{\nu}}=\dfrac{g_{a}}{g_{\nu}}\left(\dfrac{T_{\text{a}}}{T_{\nu}}\right)^4\\
&\simeq& 0.027\,g_{a} \left(\dfrac{106.75}{g_{*\text{s}} (T_\text{dec})}\right)^{4/3},
\end{eqnarray}
where $T_\nu$ is the neutrino temperature, $g_{i}$ is the number of internal degrees of freedom in a given species (times 7/8 for a fermion), so that $g_{a}=1$ for an axion, while $g_{* \text{s}}$ is the total number of entropy degrees of freedom in the plasma, and $T_\text{dec}$ is the temperature at which the axion decouples from the primordial plasma. The smaller is $T_\text{dec}$ the larger is the contribution to $\Delta N_{\text{eff}}$. For this reason, in this work we will focus only on axion interactions which can lead to the lowest decoupling temperatures, around or below the QCD PT~(see instead~\cite{Turner:1986tb,Masso:2002np,Graf:2010tv,Salvio:2013iaa, Brust:2013xpv,Baumann:2016wac,Ferreira:2018vjj, DEramo:2018vss, Arias-Aragon:2020qtn} for decoupling above the QCD PT), thus to the largest value of $\Delta N_{\text{eff}}$. These are the interactions with pions and leptons, which we now separately introduce.


\subsection{Interactions with pions}\label{sec:api}

\begin{figure}[t]
	\centering
	\includegraphics[width=0.48\textwidth]{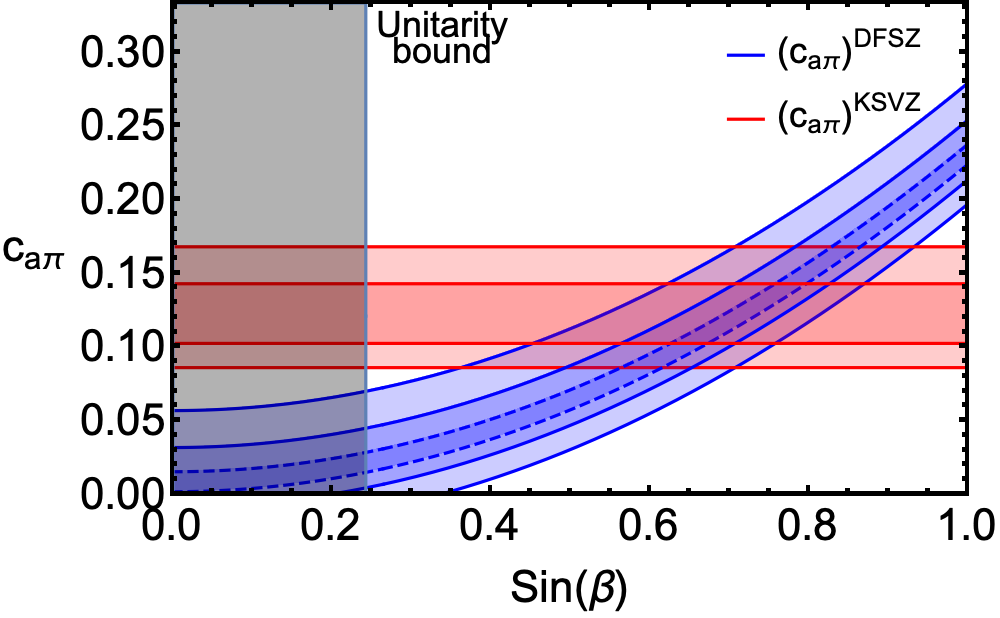}
	\caption{\small The axion-pion coupling $c_{a\pi}$ for the KSVZ (red) and DFSZ (blue) models. The bands corresponds to varying the ratio $m_u/m_d$ within the $1\sigma$ (dark) and $2\sigma$ (light) uncertainty~\cite{Zyla:2020zbs}. The band bordered by the dashed lines corresponds instead to the central value and $2\sigma$ uncertainty of Ref. \cite{Gorghetto:2018ocs}. The gray band is the region excluded by unitarity bounds \cite{DiLuzio:2020wdo}. \label{fig:capi}}
\end{figure}

Below the confinement scale the axion couples to pions\footnote{Interactions with nucleons are also present but do not lead to a significant thermal production of axions, due to the small density of nucleons in the thermal plasma.} through the Lagrangian:
\begin{equation}
\label{eq:pionlag}
\mathcal{L}_{a\pi} =
\frac{c_{a\pi}}{f_\pi} 
\frac{\partial_\mu a}{f}   \left[2 \partial^\mu \pi^0 \pi^+ \pi^- - \pi_0 \left(\partial^\mu \pi^+ \pi^- 
- \pi^+ \partial^\mu \pi^-\right) \right] ,
\end{equation}
where (see e.g.~\cite{DiLuzio:2020wdo})
\beq 
\label{eq:Capidef}
c_{a\pi} =  - \frac{1}{3} 
(c_u - c_d) = - \frac{1}{3} 
\( c^0_u - c^0_d - \frac{1-z}{1+z}  \) \, .
\eeq
Depending on the PQ breaking model and taking the uncertainties on $z$ into account \cite{Zyla:2020zbs}, we have:
\begin{align}
\textbf{KSVZ}&: \quad c_{a\pi} \simeq 0.12_{-0.018}^{+0.023} \, ,\\
\textbf{DFSZ}&: \quad c_{a\pi} \simeq 0.12_{-0.018}^{+0.023} - \frac{1}{9}\cos(2\beta) \, . \label{DFSZ}
\end{align}
The values of $c_{a\pi}$ as a function of $\sin\beta$ are shown in Fig.~\ref{fig:capi} for the two models. We have included constraints on $\sin\beta$, from tree-level unitarity of fermion scattering~\cite{Bjorkeroth:2019jtx, DiLuzio:2020wdo}, i.e.~$\tan \beta \in  \left[ 0.25, 170\right]$ (the upper bound on $\sin\beta$ is not visible in the plot, since it corresponds to $\beta\simeq \pi/2$). Thanks to the additional dependence on $\sin\beta$, the axion-pion coupling in the DFSZ model can be significantly suppressed compared to its value in the KSVZ model when $\sin\beta$ is small, and can actually even vanish when the $2\sigma$ uncertainty on $z$ from~\cite{Zyla:2020zbs} is taken into account. Using instead the most recent lattice results~\cite{Gorghetto:2018ocs} yields $c_{a\pi}\gtrsim 0.014$ at $2\sigma$. On the other hand, when $\beta\approx \pi/2$ the axion-pion coupling is about twice as large as in the KSVZ model. 

According to the Lagrangian~\eqref{eq:pionlag}, axions can be thermally produced by scattering processes: $\pi^0\pi^{\pm }\rightarrow a\pi^{\pm}$ and $\pi^+\pi^{-}\rightarrow a\pi^0$. The thermally averaged rates for these processes were first computed by~\cite{Chang:1993gm} and updated by~\cite{Hannestad:2005df}, which found
\begin{equation}
	\label{eq:pionrate}
	\Gamma_{\pi\pi\rightarrow a\pi} = A\left(\frac{c_{a\pi}}{f_{\pi} f}\right)^2 T^5 h(x_{\pi}),
\end{equation}
where $x_{\pi} \equiv m_{\pi}/T$, $A=0.215$ and $h(x)$ is a numerical function normalized to $h(0)=1$ that decays exponentially below $T=m_{\pi}$ (see Appendix~\ref{sec:rates}).
The rate above leads to thermal decoupling below the QCD phase transition for $m_a\gtrsim 0.1~\text{eV}$, whenever $c_{a\pi}\gtrsim 0.1$. For $c_{a\pi}\ll 0.1$, \eqref{eq:pionrate} predicts a decoupling temperature above the QCD phase transition for $m_a\lesssim 1~\text{eV}$ and so, in this case,~\eqref{eq:pionrate} cannot be used to reliably compute the total thermal production of axions from pion scatterings. We discuss the associated uncertainties in the calculation of $\Delta N_{\text{eff}}$ below and report detailed calculations of the decoupling temperature in Appendix~\ref{sec:rates}.

To summarize, the dependence on $\sin\beta$ implies that thermal production of axions from pion scattering in the DFSZ model can be significantly suppressed or enhanced compared to the KSVZ model. This leads to important differences with respect to previous work, which assumed KSVZ scattering rates (see e.g.~\cite{DiValentino:2015wba, Millea:2020xxp, Giare:2020vzo}).

\subsection{Interactions with leptons}\label{sec:alep}

\noindent We now turn to the axion couplings to leptons ($\ell$). In the DFSZ scenario these couplings are universal and arise at tree level. However, two submodels exist, depending on whether the coupling to leptons is identified with the coupling to down or up type quarks:
\begin{align} \label{DFSZI}
	\textbf{DFSZ-I}&: \quad c_\ell =c_d^0 \, \\
	\textbf{DFSZ-II}&: \quad c_\ell=-c_u^0 \, \, , \label{DFSZII}
\end{align}
where $c_{u,d}^0$ are given in \eqref{eq:DFSZ, cu and cd}. By means of \eqref{eq:Capidef}, one can relate the axion-lepton couplings above to the coupling to pions. It is then easy to see that in the DFSZ-I case $c_\ell$ is small when $c_{a\pi}$ is small, whereas in the DFSZ-II $c_\ell$ increases as $c_{a\pi}$ decreases. 

\begin{figure}
	\centering
     \includegraphics[width=1\linewidth]{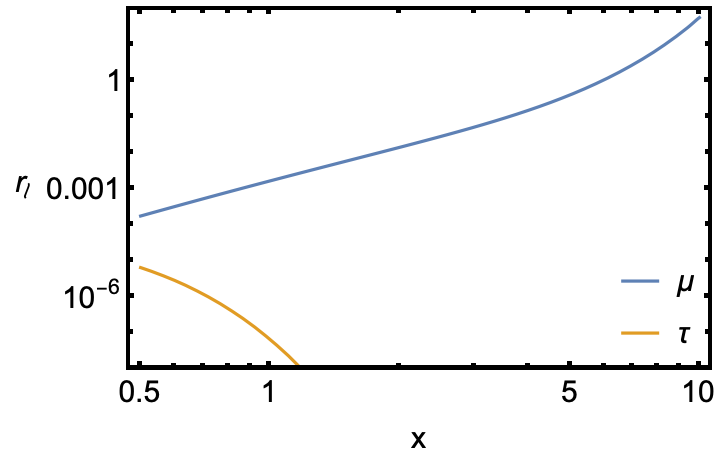}
	\caption{\small Normalized ratio of axion production rates from lepton and pion scatterings as a function of $x=m_{\pi}/T$, according to \eqref{eq:ratio} (see Appendix~\ref{sec:rates} for details).} 
	\label{fig:ratio}
\end{figure}

In the DFSZ model the axion can then be produced at tree level through scatterings involving a photon: $\ell^\pm \gamma \rightarrow \ell^\pm  a, \, \ell^+ \ell^-   \rightarrow  \gamma a$. The total thermally averaged rate is given by~\cite{DEramo:2018vss}
\begin{equation}
\label{eq:leptonrate}
\Gamma_{a\ell \rightarrow \ell \gamma}= B~c_\ell^2~\left(\frac{m_\ell}{f}\right)^2~T~g (x_\ell),
\end{equation}
where $x_\ell\equiv m_\ell/T$, $B\approx 1.3\cdot 10^{-3}$ and $g_\ell$ is a function normalized to $g_\ell(0)=1$ that decays exponentially below $T=m_\ell$ (see Appendix~\ref{sec:rates}). We thus have 
\begin{equation}
\label{eq:ratio}
\frac{\Gamma_{a\ell \rightarrow \ell \gamma}}{\Gamma_{\pi\pi\rightarrow a\pi}}=\left(\frac{c_\ell}{c_{a\pi}}\right)^2~r_\ell\left(\frac{m_\ell}{T},\frac{m_\pi}{T}\right).
\end{equation}
The function $r_\ell$ is shown in Fig.~\ref{fig:ratio} for $\ell = \tau, \mu$. Note that $r_{\tau}\ll r_{\mu}$ and $r_{\mu}\ll 1$ for temperatures below the QCD phase transition. At temperatures $T\lesssim m_{\pi}/8$, $r_{\mu}\gtrsim 1$. However, this is relevant only when the axion does not thermalize at higher temperatures due to pion scatterings, which only occurs for $f\lesssim 10^5~\text{GeV}$ (see Fig.~\ref{fig:tdec} in Sec.~\ref{sec:rates}). This region is strongly disfavored by stellar cooling (see e.g.~\cite{DiLuzio:2020wdo}) and by laboratory searches~\cite{Akerib:2017uem, Aprile:2020tmw}.

Thus, we conclude that production from leptons is negligible compared to production from pions for the range of $f$ of interest, unless $c_{\ell}\gg c_{a\pi}$. This latter possibility occurs in the DFSZ-II case, when $\sin\beta\ll 1$. We then anticipate that in this case axions will be mainly produced by muon scatterings when $c_{a\pi}\rightarrow 0$. At larger decoupling temperatures (thus smaller values of $\Delta N_{\text{eff}}$ according to \eqref{eq:deltaneff}) the contribution from taus can also be relevant. In the DFSZ-I case, axion production from pion scatterings dominates over the production from leptons, as long as $f\gtrsim 10^5~\text{GeV}$.

\subsection{Laboratory and Astrophysical Hints}

\begin{figure*}
	\centering
	\includegraphics[width=1\textwidth]{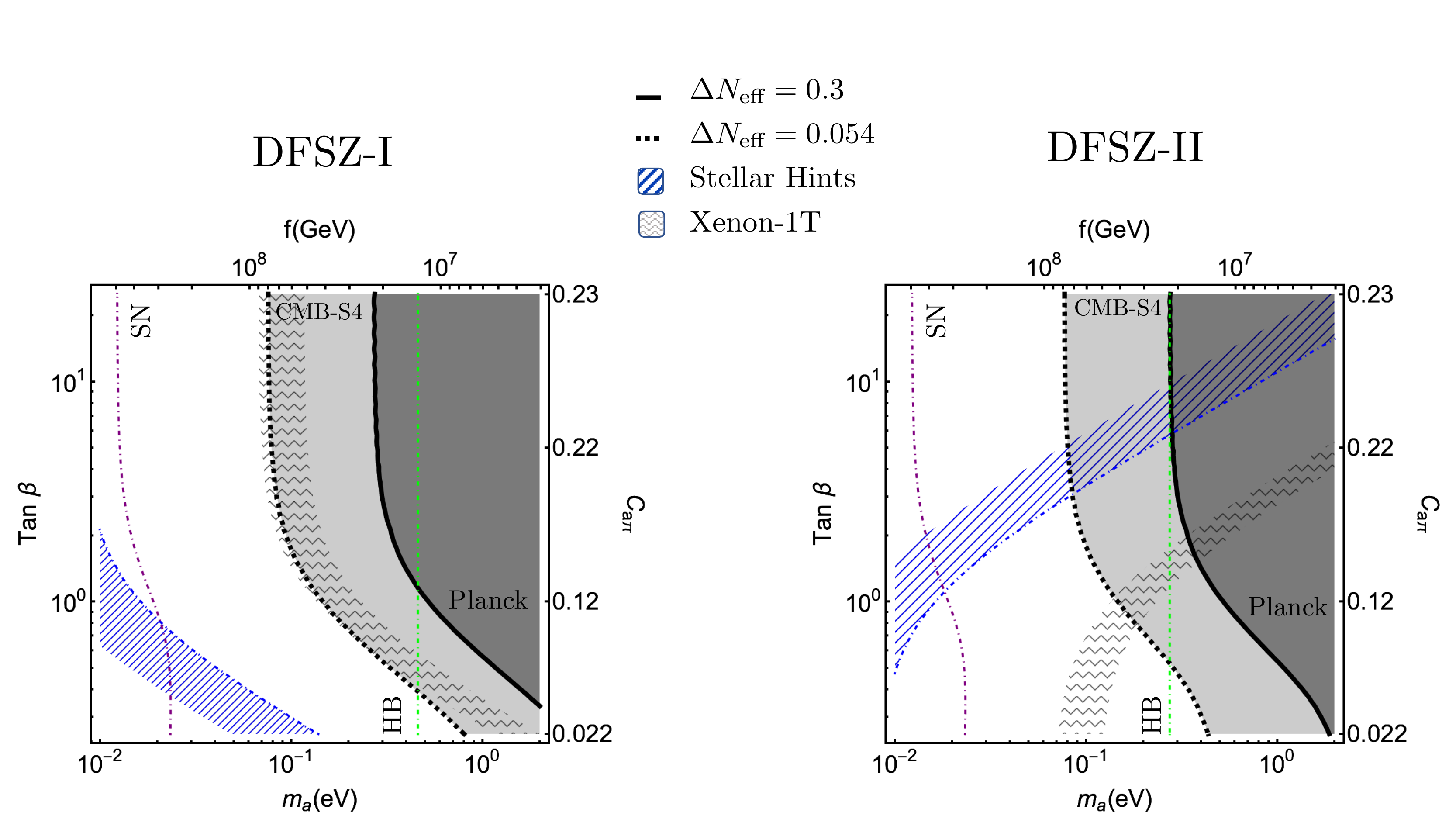}
\caption{Parameter region compatible with Xenon-1T excess (curvy gray) and stellar hints (hatched blue) for the DFSZ-I (left) and DFSZ-II (right) axion. Regions where $\Delta N_{\text{eff}}$ is large enough to be probed at $2\sigma$ by Planck18 and CMB-S4 are shaded in dark and light gray, respectively. The regions to the right of the dot-dashed blue (green, purple) line are disfavored by white dwarf (horizontal branch stars, supernovae 1987A) observations \cite{Giannotti:2017hny,Ayala:2014pea,Carenza:2019pxu}.}
\label{fig:joined}
\end{figure*}

Some recent laboratory and astrophysical observations may favor QCD axion models with significant axion-electron interactions, such as the DFSZ model. Here we consider: the X1T excess in electron-recoil events~\cite{Aprile:2020tmw} and observations of faster-than-expected cooling rates of white dwarfs~(see e.g.~\cite{Isern:2018uce, Corsico:2019nmr}) and red giant stars (see~\cite{Viaux:2013lha, Straniero:2018fbv, Straniero:2020iyi}). Both observations can be interpreted in terms of an axion particle, with suggested coupling to electrons\footnote{In the X1T case, it is in principle possible that the excess is driven by axion-photon and/or axion-nucleon couplings, in addition to the axion-electron coupling. This possibility can be realized only in the KSVZ and in the DFSZ-II model, in the region allowed by unitarity. We discard it here, since it would require $m_a\gtrsim 5~\text{eV}$ for the DFSZ-II model, which as we will see is strongly excluded by cosmology.}, typically written in terms of $g_{ae} \equiv m_e c_e/f$, given (at 1$\sigma$) by\footnote{The X1T 1-$\sigma$ confidence region was obtained by averaging the 90\%C.L. regions in each of the first two plots in figure 8 of~\cite{Aprile:2020tmw}, in the limit of vanishing photon and nucleon coupling, respectively.} 
\begin{eqnarray}
	&\textbf{Xenon-1T~\cite{Aprile:2020tmw}:} \quad	g_{ae}  \simeq&2.6^{+0.7}_{-0.7}\times 10^{-12} \label{eq:hintxenon} \, , \\
	&\textbf{Star Cooling~\cite{Giannotti:2017hny}:} \quad g_{ae} \simeq& 1.6^{+0.29}_{-0.34} \times 10^{-13} \,. \label{eq:hintastro}
\end{eqnarray}

These hints are mutually incompatible and further observations exist~(see e.g.~\cite{DiLuzio:2020wdo} for a review), which constrain the interpretation of both the aforementioned observations in terms of an axion particle. For this reason, we will perform separate analyses in this work, both with and without taking the hints above into account. In the KSVZ model the axion does not couple to leptons at tree-level, therefore $m_{a}\gtrsim 4~\text{eV}$ ($m_{a}\gtrsim 46~\text{eV}$) is required to explain the stellar (X1T) hints. Such large axion masses in the KSVZ model have already been excluded by cosmological data~\cite{Hannestad:2005df, DiValentino:2015wba} and independently by astrophysical observations (see e.g.~\cite{DiLuzio:2020wdo} for a review).
For this reason, we focus on the DFSZ axion scenario in this work. The region of axion masses and $\tan \beta$ compatible with either the stellar cooling or the X1T hints within the unitarity bounds on $\tan \beta$ are shown in Fig.~\ref{fig:joined}.

\section{Contribution to $\Delta N_{\text{eff}}$}
\label{sec:deltaneff}

The abundance of thermally produced axions affects the epoch of recombination and is thus probed by CMB observations. This is conventionally quantified by the parameter~$\Delta N_{\text{eff}}$~\eqref{eq:deltaneff}, which can be accurately computed as (see e.g.~\cite{DEramo:2018vss})
\begin{equation}
	\label{eq:deltan2}
	\Delta N_{\text{eff}} \simeq 74.85 \, (Y^{\infty}_a)^{4/3},
\end{equation}
where $Y_a\equiv n_a/s$ is the axion abundance, $n_a$ the axion number density and $s$ the total entropy density. The superscript $\infty$ means that the quantity is evaluated at asymptotically small temperatures, since after decoupling $Y_a$ is a conserved quantity. This way, one also takes into account axion production even when the axion is not thermalized. The axion abundance $Y_a$ can be computed by solving the Boltzmann equation
\begin{equation}
	\label{eq:boltzmann}
	s H x \frac{dY_a}{dx}=\left(1-\frac{1}{3}\frac{d\ln g_{\star, s}}{d\ln x}\right)n_a^{\text{eq}} \, \sum_{i} \Gamma_{i} \left(1-\frac{Y_a}{Y_a^{\text{eq}}}\right),
\end{equation}
where  $x=m_\pi/T$, $n^{\text{eq}}_a=\zeta(3)T^3/\pi^2$, $Y^\text{eq}_a=n^{\text{eq}}/s$ are the equilibrium values for the axion number density and abundance, and $\sum_i \Gamma_{i}$ sums over the thermally averaged scattering rates of the different production channels. Due to the effects of the QCD phase transition, the behavior of $g_{* s}$ has to be extracted from lattice computations (we use the most recent results from~\cite{Borsanyi:2016ksw}). 

In this work we focus on the contribution to $\Delta N_{\text{eff}}$ from pion~\eqref{eq:pionrate}, muon and tau~\eqref{eq:leptonrate} scatterings. These are the relevant processes for thermalization around or below the QCD phase transition, thus leading to a value of $\Delta N_{\text{eff}}$ which can be probed by current data.\footnote{See instead~\cite{Turner:1986tb,Masso:2002np,Graf:2010tv,Salvio:2013iaa,Arias-Aragon:2020shv} for thermalization above and across the EW PT and~\cite{Brust:2013xpv,Baumann:2016wac,Ferreira:2018vjj, DEramo:2018vss} for thermalization below the EW PT but excluding pions (see also \cite{Arias-Aragon:2020qtn} for the possible correlations with the X1T excess).} 
We solve the Boltzmann equation \eqref{eq:boltzmann} from $T_i = 100~m_\pi$ to $T_f= 1~\text{MeV}$, setting the initial axion abundance to zero. 
The contribution from pions relies on the thermally averaged pion scattering rate \eqref{eq:pionrate} which ceases to be reliable at temperatures larger than roughly $160~\text{MeV}$~(see e.g. the behavior of $g_{\star, s}$ in~\cite{Borsanyi:2016ksw}), the precise value requiring a dedicated analysis beyond the scope of this paper. For definiteness, in this work we use the latest determination of the QCD critical temperature from lattice calculations, i.e.~$T_{\text{QCD}}\simeq 158$~MeV~\cite{Borsanyi:2020fev}, and include pions in the Boltzmann equation only below this temperature (see Appendix~\ref{sec:boltzmann} for a discussion on how our results are affected by a different choice of this cutoff temperature). On the other hand, when $T_{\text{dec}}>T_{\text{QCD}}$, our computation can only provide a lower bound on $\Delta N_{\text{eff}}$, since it is in principle possible that a sizable contribution arises from scatterings with pions, or heavier mesons and baryons, during the QCD phase transition. Nonetheless, an upper bound on such a contribution can be estimated assuming $T_{\text{dec}}=T_{\text{QCD}}$, since decoupling at higher temperatures would give a smaller $\Delta N_{\text{eff}}$ according to~\eqref{eq:deltaneff}. Regarding the contribution from leptons, it is only relevant for the DFSZ-II model at small values of $\tan\beta$, as discussed in Sec.~\ref{sec:alep}. 

Contours of $\Delta N_{\text{eff}}$ in the parameter space of the DFSZ model are shown in Fig.~\ref{fig:joined}, obtained by also fixing $z=0.47$ and $T_{\text{QCD}}\simeq 158$~MeV (see Appendix~\ref{sec:boltzmann} for further results). We focus on values that can be constrained at $2\sigma$ by current (Planck) and future (CMB-S4~\cite{Abazajian:2016yjj}) CMB datasets. We find that the existence of a DFSZ axion with $m_{a}\gtrsim 0.3~$eV can be probed using Planck data at $2\sigma$ significance; this is true for both the DFSZ-I and DFSZ-II submodels, when $\tan\beta\gtrsim 1$. At smaller values of $\tan\beta$, we see that $\Delta N_{\text{eff}}$ is suppressed, as expected from our discussion in Secs.~\ref{sec:api} and \ref{sec:alep}, and we expect constraints on the axion mass to be significantly relaxed (indeed, when taking uncertainties on $z$ into account, axion production can be significantly shut off). In the DFSZ-II submodel, axion production from lepton scattering remains relevant even at small values of $\tan\beta$, as discussed in Sec.~\ref{sec:alep}, thus reducing the possibility to relax constraints. When comparing with the constraints coming from astrophysical observations, we see that current cosmological data is competitive with the horizontal branch stars bound \cite{Ayala:2014pea} and it can be more constraining than white dwarf cooling observations \cite{Giannotti:2017hny} in the DFSZ-II case for large $c_{a\pi}$.

Fig.~\ref{fig:joined} also shows that current hints for a DFSZ axion can be independently investigated by means of cosmological data. This is especially true for the DFSZ-II scenario, where both the stellar cooling and X1T hints lead to observable values of $\Delta N_{\text{eff}}$ for $m\gtrsim 0.1~\text{eV}$. For the DFSZ-I scenario, only the region suggested by the X1T hint leads to an observable value of $\Delta N_{\text{eff}}$. Interestingly, most of this region can be probed at $2\sigma$ by CMB-S4.

\section{Datasets and Results}
\label{sec:data}

We now turn to a quantitative assessment of the DFSZ axion scenario in light of the latest cosmological datasets. To this aim, we use the Boltzmann code {\tt{CLASS}}~\cite{Lesgourgues:2011re, Blas:2011rf}, which we have modified to include the thermally produced QCD axion as an extra light species beyond the SM neutrinos, in complete analogy with the case of an extra neutrino species which is already included in {\tt{CLASS}}.
In particular, we have implemented the following changes:
\begin{itemize}
\item[$1$] We used a Bose-Einstein distribution with one degree of freedom to describe the thermally produced axions, rather than the Fermi-Dirac distribution which is implemented in {\tt{CLASS}} for extra neutrino species.
\item[$2$] We assigned a temperature to the extra species, given by $T_a = T_{\nu}\, \left(\frac{7}{4} \Delta N_{\text{eff}} \right)^{1/4}$, as dictated by \eqref{eq:deltaneff}.
\end{itemize}
With these modifications, we obtain a cosmological model with six $\Lambda$CDM parameters plus two additional parameters: $c_{a\pi}$ (or alternatively $\sin\beta$) and $m_a$.  
We then use {\tt Monte Python}~\cite{Audren:2012wb, Brinckmann:2018cvx}, in its version $3.3.2$, to perform a Markov chain Monte Carlo analysis of this model. We model neutrinos, using the standard treatment of the Planck collaboration, as two massless 
and one massive species with $m_\nu = 0.06$ eV and $T_\nu=0.71611~\text{eV}$.\footnote{We checked that adding $\sum m_{\nu}$ as a free parameter in our runs does not significantly affect the upper bounds on $m_a$ presented here.} 
We obtain constraints on $m_a$ by means of the following two combinations of cosmological datasets:
\begin{itemize}
\item[a)]{\textbf{Planck 18 + BAO}: Planck 2018 high-$\ell$ and low-$\ell$ TT, TE, EE and lensing data~\cite{Aghanim:2019ame}, plus BAO measurements from 6dFGS at $z = 0.106$~\cite{Beutler:2011hx}, from the MGS galaxy sample of SDSS at $z = 0.15$~\cite{Ross:2014qpa}, and
  from the CMASS and LOWZ galaxy samples of BOSS DR12 at $z = 0.38$, $0.51$, and $0.61$~\cite{Alam:2016hwk}.}
\item[b)]{\textbf{Planck 18 + BAO + Pantheon + SH$_0$ES}: same as above plus SH$_{0}$ES 2019 measurement of the present day Hubble rate $H_0 = 74.03\pm 1.42$ km/s/Mpc~\cite{Riess:2019cxk}, and the Pantheon supernovae dataset~\cite{Scolnic:2017caz}.}
\end{itemize}
We first focus on general constraints, obtained using only the datasets a) and b) above, which in our figures are referred to as ``w/o SN + H$_0$'' and ``w/ SN + H$_0$'' respectively. Note that 
we improve the analysis compared to previous related work~\cite{Hannestad:2005df, Archidiacono:2013cha, DiValentino:2015wba, Millea:2020xxp} by solving the Boltzmann equation to obtain the axion abundance rather than using approximate estimations of the decoupling temperature.

\begin{figure}
	\centering
	\includegraphics[width=\linewidth]{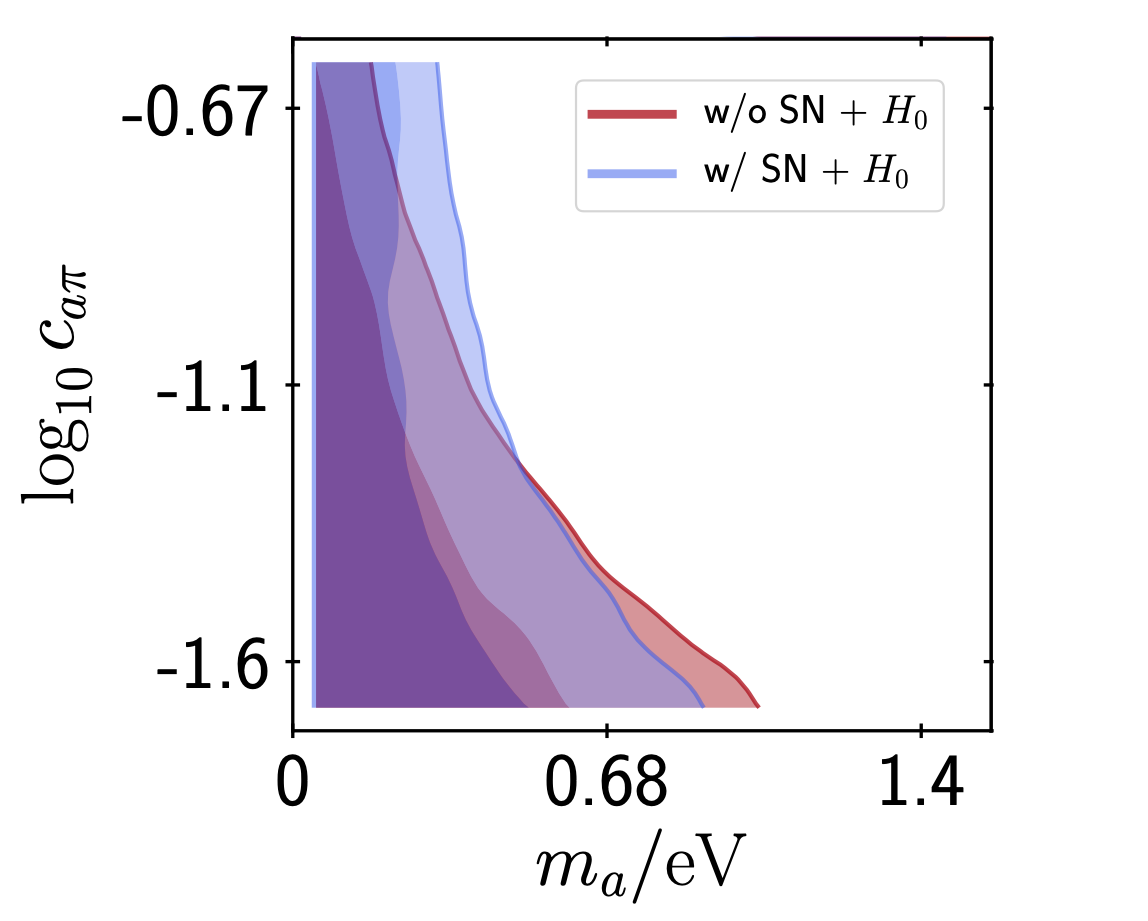}
	\caption{\small Constraints on the axion coupling to pions $c_{a\pi}$ and on the axion mass $m_a$ when axion production via leptons can be neglected. This is the case for the the DFSZ-I model at any value of $c_{a\pi}$, whereas in the DFSZ-II model the constraints shown in the figure are reliable only for $c_{a\pi} \gtrsim  {\cal O}(0.1)$.} 
	\label{fig:dfszI}
\end{figure}

As discussed in Sec.~\ref{sec:alep}, in the DFSZ-I model thermal axion production is dominated by pion scatterings independently of the value of the axion-pion coupling. In this case, the contribution to $\Delta N_{\text{eff}}$ has a simple dependence on $c_{a\pi}$, and thus we choose the latter as an independent parameter in our MonteCarlo analysis, together with $m_a$ (our results then only depend weakly on $z$ through the axion mass~\eqref{eq:axmass}, therefore we fix $z=0.47$ in these runs). We show the resulting constraints on $m_a$ and $c_{a\pi}$ in Fig.~\ref{fig:dfszI}. We use a logarithmic prior on $c_{a\pi}$: $\ln c_{a\pi}=[-4, -1.27]$ to encompass both small and large values of $c_{a\pi}$, see Fig.~\ref{fig:capi}.

\begin{figure*}[t]
	\centering
	\includegraphics[width=0.8\linewidth]{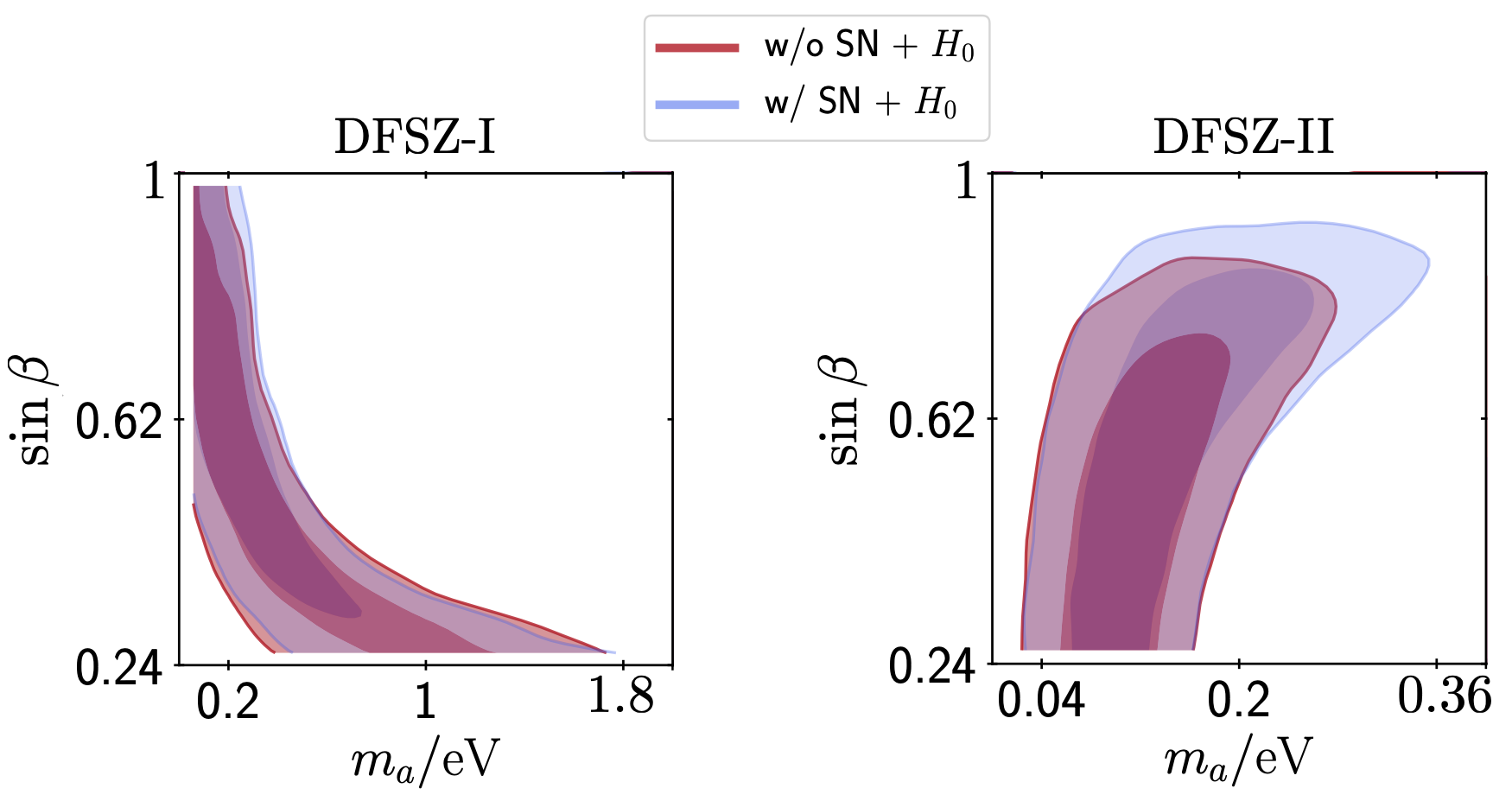}
	\caption{\small Constraints on the axion mass and the angle $\beta$ between the two Higgs doublets in the DFSZ model, obtained by combining cosmological data Planck18+BAO (+Pantheon+SH$_0$ES) with the Xenon-1T hint~\eqref{eq:hintxenon}. \emph{Left:} DFSZ-I model,  \emph{Right:} DFSZ-II model. Dark (light) shaded contours correspond to 1(2)$\sigma$ regions.} 
	\label{fig:xenon}
\end{figure*}

One can appreciate that the cosmological bound on $m_a$ becomes stronger as $c_{a\pi}$ increases whereas constraints are significantly relaxed as $c_{a\pi}$ decreases. In order to derive precise bounds, we perform a dedicated analysis of two representative values of $c_{a\pi}= 0.225, 0.0225$, roughly corresponding to the maximal and minimal central values of $c_{a\pi}$ (see~Fig.~\ref{fig:capi}). The corresponding upper bounds on $m_a$ at $95\%$ C.L. are reported in Table~\ref{tab:constraints} for our two datasets. Smaller values of $c_{a\pi}$ are possible at $2\sigma$ when using the uncertainty on $z$ from \cite{Zyla:2020zbs} (see Fig.~\ref{fig:capi}) and as $c_{a\pi}\rightarrow 0$ the constraint on $m_a$ is fully removed.

A separate analysis is required for the DFSZ-II model, since thermal production from leptons is relevant in this case for small values of $c_{a\pi}$, as discussed in Sec.~\ref{sec:alep}. By including both leptons and pion scattering rates in the Boltzmann equation and implementing the results in  {\tt{CLASS}} as described above, we obtain the upper bound reported in Table~\ref{tab:constraints} for the same representative values of $c_{a\pi}$.  As expected, when $c_{a\pi}$ is small, $c_{a\pi}= 0.0225$, we find stronger constraints than in the DFSZ-I model, driven by the extra contribution from leptons. On the other hand, when $c_{a\pi}$ is ${\cal O}(0.1)$ the contribution from leptons is negligible and the bound is the same as in the DFSZ-I model.

We now turn to the results obtained by combining cosmological datasets with either laboratory or astrophysical hints, which we implement by including a Gaussian likelihood on $g_{ae}$ in our runs, according to~\eqref{eq:hintxenon} and~\eqref{eq:hintastro}. In these cases, thermal axion production from leptons is always negligible compared to pions. For convenience, we trade the parameter $c_{a\pi}$ for $\sin\beta$ (or $\tan\beta$) and impose unitarity constraints on the latter parameters by choosing flat priors on $\sin\beta=[0.24, 1]$. In order to take into account the possibility that $c_{a\pi}\rightarrow 0$ as $z$ varies within its experimental uncertainties, we now keep $z$ as an extra parameter, which we constrain by including a Gaussian likelihood according to~\cite{Zyla:2020zbs}.

\begin{table}
\vspace{0.5cm}
\centering
{\begin{tabular}{|c | c |}
\hline
DFSZ-I & Planck 18+BAO (+SN+$H_0$)\\
\hline
$c_{a\pi} = 0.225$ & $m_a\leq 0.20~(0.29)~\text{eV}$ \\
$c_{a\pi} = 0.0225$ & $m_a\leq 0.84~(0.82)~\text{eV}$ \\
\hline
\end{tabular}}
{\begin{tabular}{|c | c |}
\hline
DFSZ-II & Planck 18+BAO (+SN+$H_0$)\\
\hline
$c_{a\pi} = 0.225$ & $m_a\leq 0.20~(0.29)~\text{eV} $ \\
$c_{a\pi} = 0.0225$ & $m_a\leq 0.60~(0.61)~\text{eV}$ \\
\hline
\end{tabular}}
\caption{\small Constraints on $m_a$ for the DFSZ-I and DFSZ-II models at $95\%$ C.L., for two representative choices of axion coupling to pions.}
\label{tab:constraints}
\end{table}
\vspace{1cm}
We start by considering the X1T hint~\eqref{eq:hintxenon}. We show the combined constraints on the fundamental parameters $m_a$ and $\sin\beta$ of the DFSZ-I and DFSZ-II models in Fig.~\ref{fig:xenon}. In the DFSZ-I case the constraints on $m_a$ are significantly relaxed at small values of $\sin\beta$, as expected from Fig.~\ref{fig:joined}. In particular, we find $m_a\lesssim 1.8~\text{eV}$ when $\sin \beta$ takes the minimal value allowed by the unitarity constraint, $\sin\beta=0.24$, with the Planck18+BAO+X1T dataset. The parameter $\sin \beta$ is, on the other hand, unconstrained. 
This is in contrast with the DFSZ-II case, where the constraint on $m_a$ is only slightly relaxed at large values of $\sin\beta$. In particular, we find $m_a\lesssim 0.27~\text{eV}$ for $\sin \beta$ close to its largest allowed value, which we find to be $\sin\beta \lesssim 0.9$ using the Planck18+BAO+X1T dataset. These differences between the two submodels can be understood from Fig.~\ref{fig:joined}: in the DFSZ-II case $\Delta N_{\text{eff}}$ increases rapidly when the mass increases, whereas in the DFSZ-I case the amount of relic axions in the X1T band is almost constant and the main cosmological effect is the fact that the axion behaves as a warm dark matter component. When using the dataset \textit{b)} we find essentially no difference for the DFSZ-I case, whereas the constraint is further relaxed in the DFSZ-II case. The SH$_0$ES measurement of $H_0$, being strongly in tension with the other datasets, prefers a larger $\Delta N_{\text{eff}}$, thereby allowing for slightly larger axion masses. 
\begin{figure*}[t]
	\centering
	\includegraphics[width=0.8\linewidth]{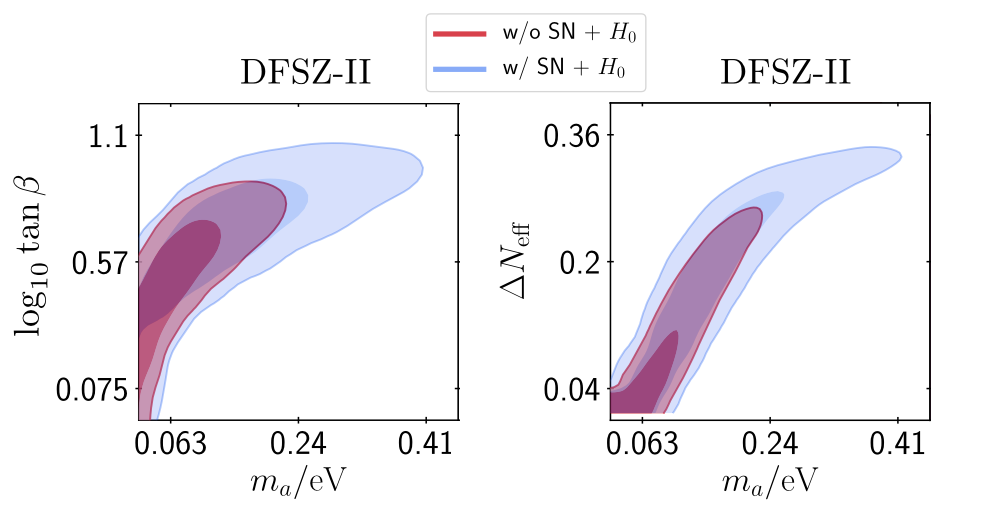}
	\caption{\small \emph{Left:} Constraints on the axion mass and the angle $\beta$ between the two Higgs doublets in the DFSZ-II model. \emph{Right:} Constraints on the axion mass and the amount of relic axions $\Delta N_\text{eff}$. Both  figures are obtained by combining cosmological data Planck18 + BAO (+Pantheon+SH$_0$ES) with the stellar cooling hint~\eqref{eq:hintastro}. Dark (light) shaded contours correspond to 1(2)$\sigma$ regions.} 
	\label{fig:stellar}
\end{figure*}
We now move to the stellar hints. We focus on the DFSZ-II case, since in the DFSZ-I scenario thermal axion production is small in the stellar hint band (see left plot in Fig.~\ref{fig:joined}). In this particular analysis, it is convenient to trade $\sin\beta$ for $\tan\beta$ to facilitate convergence and impose a logarithmic prior $\log_{10}\tan\beta=[-0.6, 2.23]$ from unitarity. We show the combined constraints on the fundamental parameters $m_a$ and $\tan\beta$ of the DFSZ-II model in Fig.~\ref{fig:stellar} (left). As expected from the discussion above, we find strong constraints, $m_a\lesssim 0.21~\text{eV}$ when $\log_{10}\tan\beta$ is close to its largest allowed value, which we find to be $\log_{10}\tan\beta \lesssim 0.85$, and a significant relaxation of these bounds when SH$_0$ES is included. We also show the combined constraints on $\Delta N_{\text{eff}}$ and $m_a$: as expected from Fig.~\ref{fig:joined} (right), a significant amount of relic axions can be produced in the stellar hint band, which then drives strong constraints on the axion mass.

To summarize, we find that current cosmological data impose strong constraints on the axion mass $m_a$ in the DFSZ-II scenario, which allow to rule out a significant fraction of the parameter space in which the QCD axion can explain the X1T excess or the stellar cooling hints. On the other hand, we find that the DFSZ-I model is significantly less constrained by current cosmological data. Next generation of CMB experiments will be able to further improve the bounds on the axion mass. In particular, they will be able to probe most of the parameter space where the DFSZ-I model can explain the X1T excess (see~Fig.~\ref{fig:joined}).

\section{Conclusions}
\label{sec:conclusions}
A sizeable thermal axion relic abundance can affect the CMB in a way similar to massive neutrinos. In this work, we used the latest cosmological data to constrain the DFSZ scenario, when axions are produced through scatterings of pions and leptons. This occurs for $m_a\gtrsim 0.1~\text{eV}$ and thus our analysis provides constraints which are independent from, and in some cases stronger than, those derived with astrophysical observations (excluding SN1987A) while relying on the arguably simpler physics of the CMB.

In contrast with KSVZ axions, in the DFSZ model the axion coupling to pions, $c_{a\pi}$, is not fixed and can be enhanced or suppressed, depending on the vevs of two Higgs doublets which characterize this class of realizations of the Peccei-Quinn mechanism. Therefore, cosmological constraints on the axion mass strongly depend on such coupling. In particular, by solving the Boltzmann equation, implementing the thermal axion in a Boltzmann code and performing a MCMC analysis, we found that: on the one hand, for maximal axion-pion couplings the axion mass is constrained to $m_a \leq 0.2$ eV by Planck and BAO data at $95\%$ C.L.; on the other hand, such constraint can be significantly alleviated when the axion-pion coupling is suppressed. In such a pionphobic limit it becomes possible to observationally distinguish between different options for the axion coupling to leptons: indeed, while in so-called DFSZ-I models this coupling vanishes simultaneously with the coupling to pions, the opposite is true in the DFSZ-II and scatterings with muons become more relevant as $c_{a\pi}$ is reduced. For instance, when $c_{a\pi}$ is suppressed by a factor of 10 from its maximal value, we find $m_a\leq 0.60 \,(0.84)$ eV at $95\%$ C.L. in the DFSZ-II (DFSZ-I) case and for even smaller values of $c_{a\pi}$ only the DFSZ-II model can be constrained by CMB observations.

Our approach is particularly well motivated in light of recent hints of a DFSZ-like QCD axion coupling to electrons, both from Xenon-1T experiment and from observations of stellar coolings. In particular, the former is in tension with astrophysical constraints and it is thus crucial to understand the extent to which current and future cosmological data can shed light on the viability of axion models to explain such signals. In the DFSZ model, the couplings to leptons is universal and related to the coupling to pions, thus the hints can potentially have direct implications for the CMB. We performed a combined analysis and found that CMB data already restricts the mass range in which the DFSZ axion addresses the Xenon-1T excess: this is particularly so for the DFSZ-II scenario, where we determined $0.07~\text{eV} \lesssim m_a \lesssim 0.3~\text{eV}$, while the upper bound is significantly weaker in the DFSZ-I case, $m_a \lesssim 1.8~\text{eV}$, due to the possible suppression of the axion-pion coupling. Interestingly, we showed that next generation CMB observations will be able to probe at $2\sigma$ most of the parameter space of the DFSZ-I model, if the latter is to address the X1T excess.
In the case of stellar hints, the reported axion-electron coupling is smaller and so we were only able to constrain the interpretation within the DFSZ-II model to the mass range $3~\text{meV} \lesssim m_a \lesssim 0.2 ~\text{eV}$. 

In this work, we have restricted the computation of thermal axion production from pions below the QCD PT, where the pion scattering rate is reliable. This implies that only the mass range $m\gtrsim 0.3\, (0.1)\text{eV}$ can be probed by current (future) CMB observations with $2\sigma$ significance. However, it is in principle possible that a further significant contribution to $\Delta N_{\text{eff}}$ arises from scatterings across the QCD PT (pions and/or heavier mesons). This would then extend the mass range that can be probed by CMB-S4 at $2\sigma$, possibly down to  masses which are closer to the SN 1987A bound, and thus provide the opportunity to firmly establish the lower bound on the QCD axion mass. Extending the computation of $\Delta N_{\text{eff}}$ across the QCD PT is thus a very interesting task for future work.

Furthermore, we have restricted our attention to the minimal particle content of QCD axion models. However, the latter may in general couple to extra fields, belonging for instance to a dark sector. This possibility may in fact be particularly well motivated in the DFSZ scenario, where solving the domain wall problem requires additional contributions to the axion potential (see~\cite{Ferrer:2018uiu} for a solution with a light dark sector). In this case, scatterings of dark sector fields may also lead to a larger value of $\Delta N_{\text{eff}}$. We leave these interesting questions for future work.

Finally, let us mention that the outlook on experimental searches for a DFSZ axion is promising: in particular the mass range considered in this work can be probed via the axion coupling to photons, by e.g. the IAXO helioscope~\cite{Irastorza:2011gs} (and/or to a lesser extent by its scaled down version BabyIAXO~\cite{Abeln:2020ywv}), and via the axion-electron coupling by next generations electron recoil detectors such as PandaX-4T~\cite{Zhang:2018xdp}, LZ~\cite{Akerib:2019fml} and XENONnT~\cite{Aprile:2020vtw}.


\section*{Acknowledgments}
We thank Luca di Luzio and Giovanni Villadoro for useful discussions. We acknowledge use of Tufts HPC research cluster. The work of A.N. is supported by the grants FPA2016-76005-C2-2-P, PID2019-108122GB-C32, "Unit of Excellence Mar\'ia de Maeztu 2020-2023" of ICCUB (CEX2019-000918-M), AGAUR2017-SGR-754. RZF acknowledges support by the Spanish Ministry MEC under grant  FPA 2017-88915-P and the Severo Ochoa excellence program of MINECO (SEV-2016- 0588). IFAE is partially funded by the CERCA program of the Generalitat de Catalunya. The work of FR is supported in part by National Science Foundation Grant No. PHY-2013953.

\appendix
\section{Scattering rates}
\label{sec:rates}

For completeness, here we summarize the results on the thermally averaged rates which have been used in this paper.

\subsection{Pions}

\begin{figure}[t]
	\centering
	\includegraphics[width=0.4\textwidth]{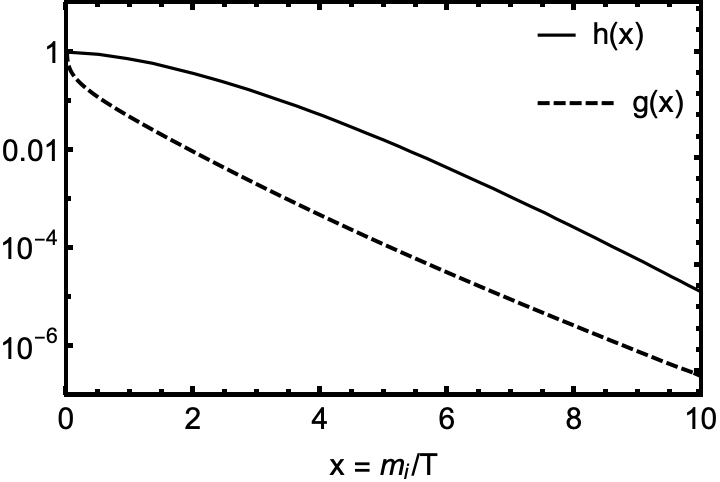}
	\caption{\small The numerical functions $h(x)$ (from~\cite{Hannestad:2005df}) and $g(x)$\label{fig:hMirizzi}.}.
\end{figure}

The thermally averaged rates for axion production from pion scatterings is given by~\cite{Chang:1993gm,Hannestad:2005df}:
\begin{equation} \label{eq: pion rate}
	\Gamma_{\pi\pi\rightarrow a\pi} = A\left(\frac{c_{a\pi}}{f_{\pi} f}\right)^2 T^5 h(x_{\pi}),
\end{equation}
where $x_{\pi} = m_{\pi}/T$, $A=0.215$ and $h(x)$ is a numerical function normalized to $h(0)=1$, shown in Fig.~\ref{fig:hMirizzi}.

\begin{figure}[t]
	\centering
	\includegraphics[width=0.9\linewidth]{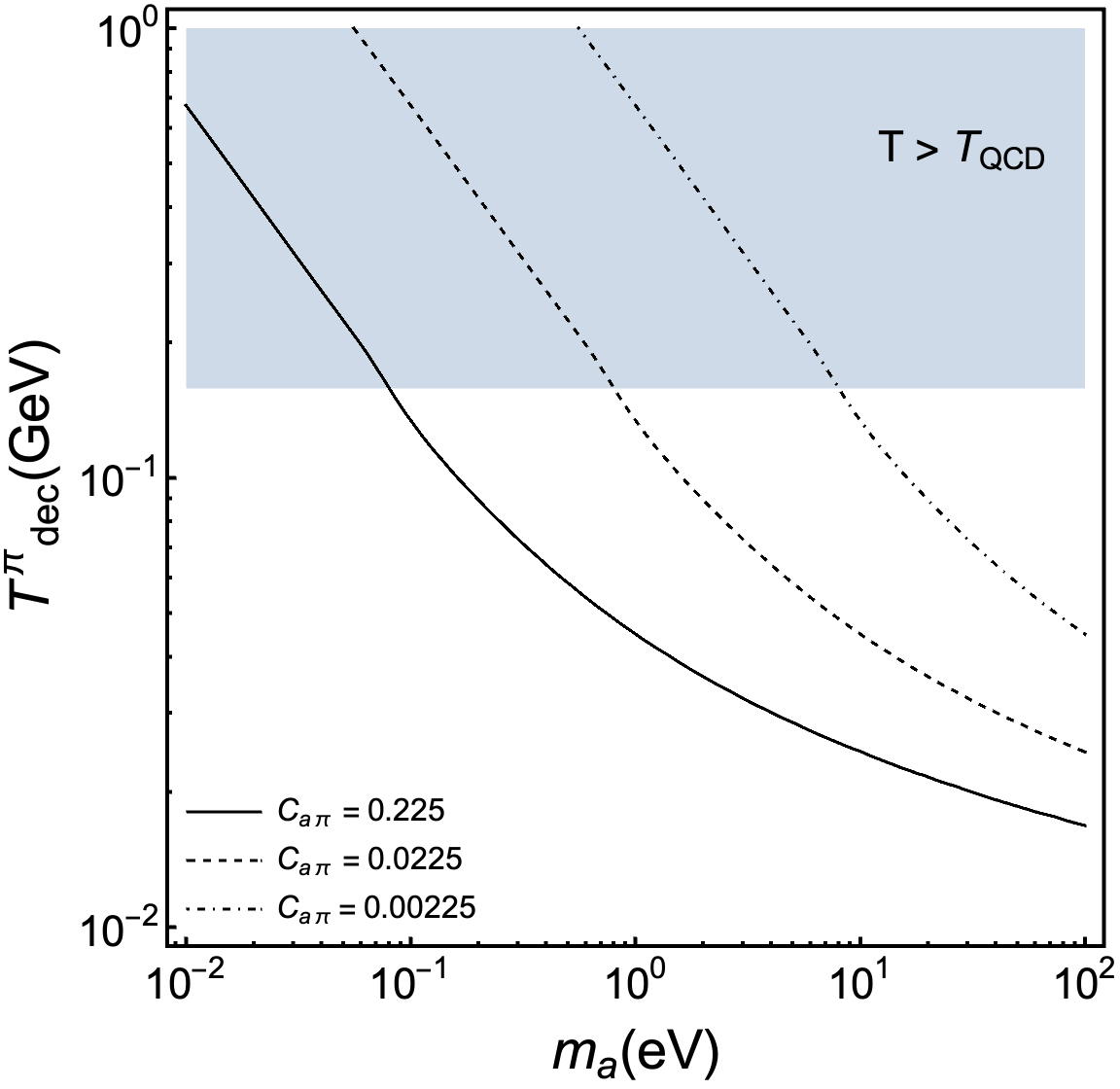}
	\caption{\small The axion decoupling temperature $T^\pi_{\text{dec}}$ as a function of the axion mass $m$ for the processes $\pi \pi\rightarrow a\pi$, according to the condition $\Gamma_{\pi\pi\rightarrow a\pi}(T^\pi_{\text{dec}})=H(T^\pi_{\text{dec}})$ and the rate \eqref{eq: pion rate}. The solid (dashed, dot-dashed) curves correspond to the choice $c_{a\pi}=0.225 \,(0.0225, 0.00225)$. The blue-shaded region corresponds to temperatures above the QCD PT, where the description in terms of pions is not valid and thus the rate \eqref{eq:pionrate} cannot be used. This is found by setting the critical temperature of the QCD PT at $T_{\text{QCD}} = 158~\text{MeV}$ \cite{Borsanyi:2020fev}. \label{fig:tdec}}
\end{figure}

The decoupling temperature can then be estimated by setting $\Gamma_{\pi\pi\rightarrow a\pi}(T^\pi_{\text{dec}})=H(T^\pi_{\text{dec}})$. The result is shown in Fig.~\ref{fig:tdec} for three representative choices of $c_{a\pi}$. The largest value of $c_{a\pi}$ in Fig.~\ref{fig:tdec} corresponds to the maximal central value of $c_{a\pi}$ in Fig.~\ref{fig:capi}.
Above the QCD phase transition, the thermal production of axions from pions cannot be reliably computed according to \eqref{eq:pionrate}. For this reason, we have shaded the region above $T= 158$ MeV, which corresponds to the most recent value of the QCD critical temperature obtained by means of lattice calculations~\cite{Borsanyi:2020fev}. We comment on the implications of the QCD phase transition on our computation of $\Delta N_{\text{eff}}$ below.

\subsection{Leptons}
The axion-lepton interaction opens up three different channels of axion production: $\ell^\pm \gamma \rightarrow \ell^\pm  a, \, \ell^+ \ell^-   \rightarrow  \gamma a$. The total thermally averaged production rate is given by the sum over the three channels and is given by (see~\cite{DEramo:2018vss} for details) 
\begin{equation}
\Gamma_{a\ell \rightarrow \ell \gamma}= B~c_\ell^2~\left(\frac{m_\ell}{f}\right)^2~T~g (x_\ell),
\end{equation}
where $B=1.3\times 10^{-3}$, $x_\ell=m_{\ell}/T$ and $g(x_\ell)$ is a numerical function normalized to $g(0)=1$, shown in Fig.~\ref{fig:hMirizzi}.

\begin{figure*}[t]
	\centering
	\includegraphics[width=0.45\linewidth]{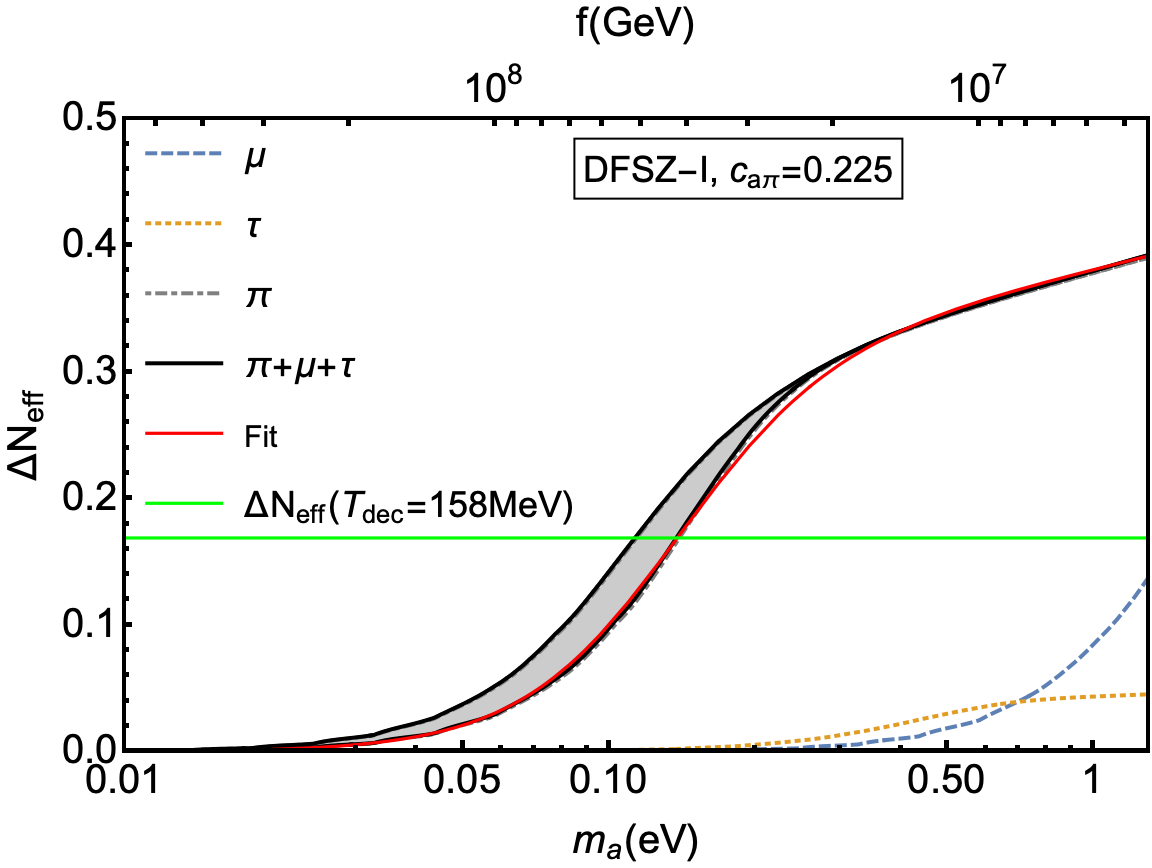}
	\, \,
	\includegraphics[width=0.45\linewidth]{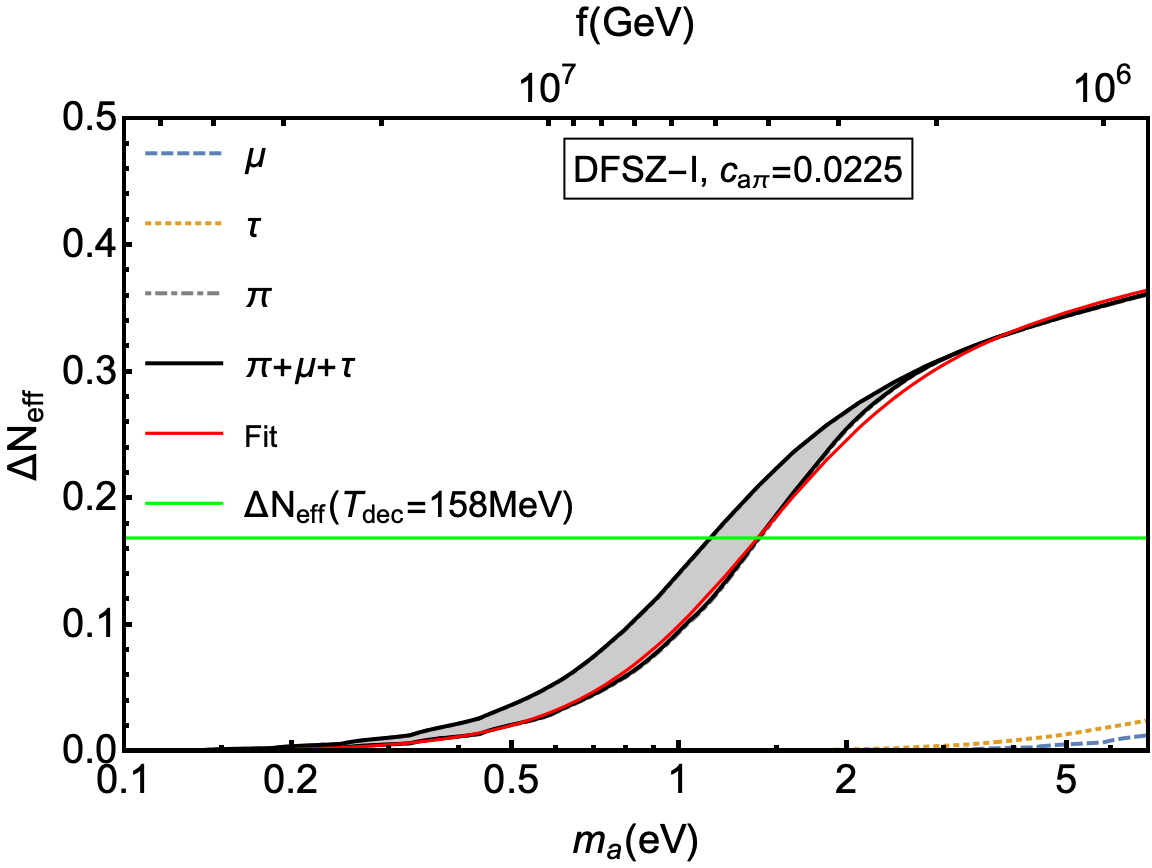}
	\caption{\small The thermal axion abundance, obtained by solving numerically the Boltzmann equation including production from pion, muon and tau scatterings, in the DFSZ-I model. We show two representative examples with $c_{a\pi}=0.225$ and $c_{a\pi}=0.0225$. Solutions obtained by considering each component separately are also shown as dashed lines in the right plot (see legends). The green line is the value of $\nef$ if the axion decouples at $T=158$ MeV. \label{fig:boltzmannI}}
\end{figure*}

\begin{figure*}[t]
	\centering
	\includegraphics[width=0.45\linewidth]{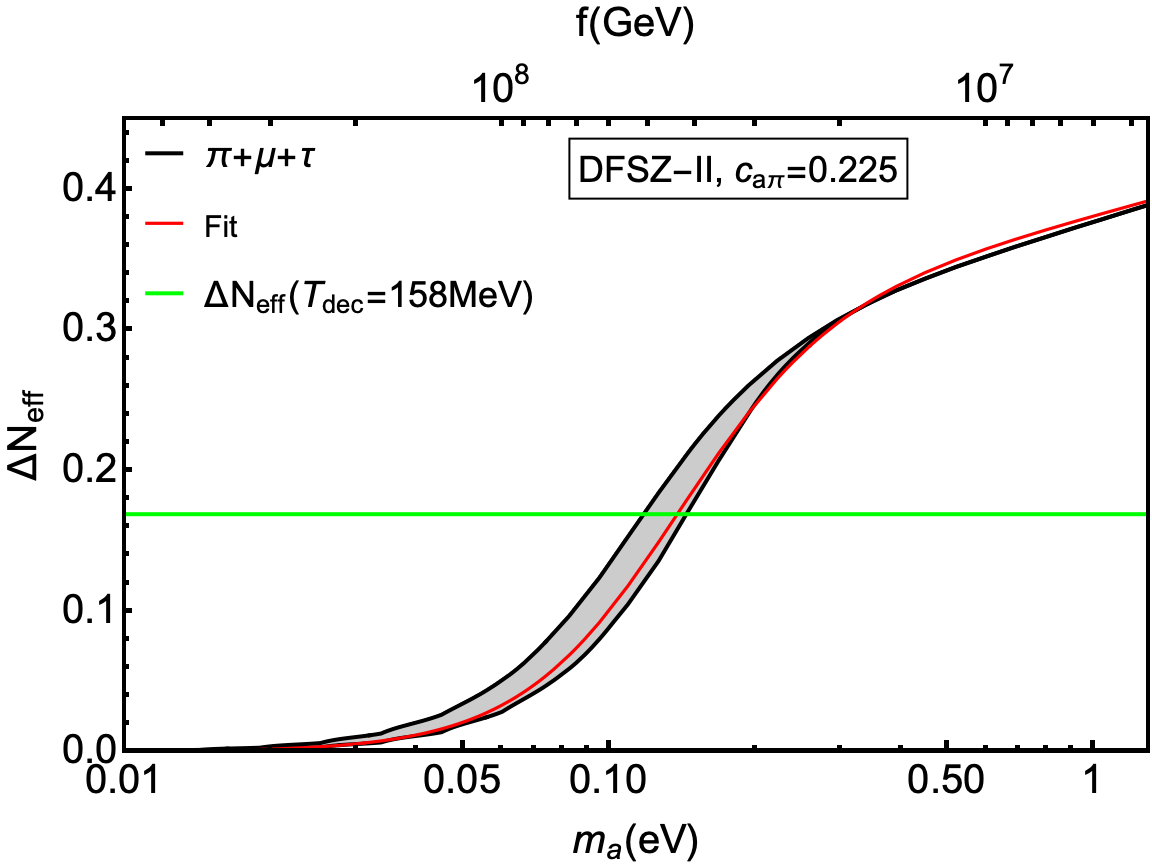}
    \, \,
	\includegraphics[width=0.45\linewidth]{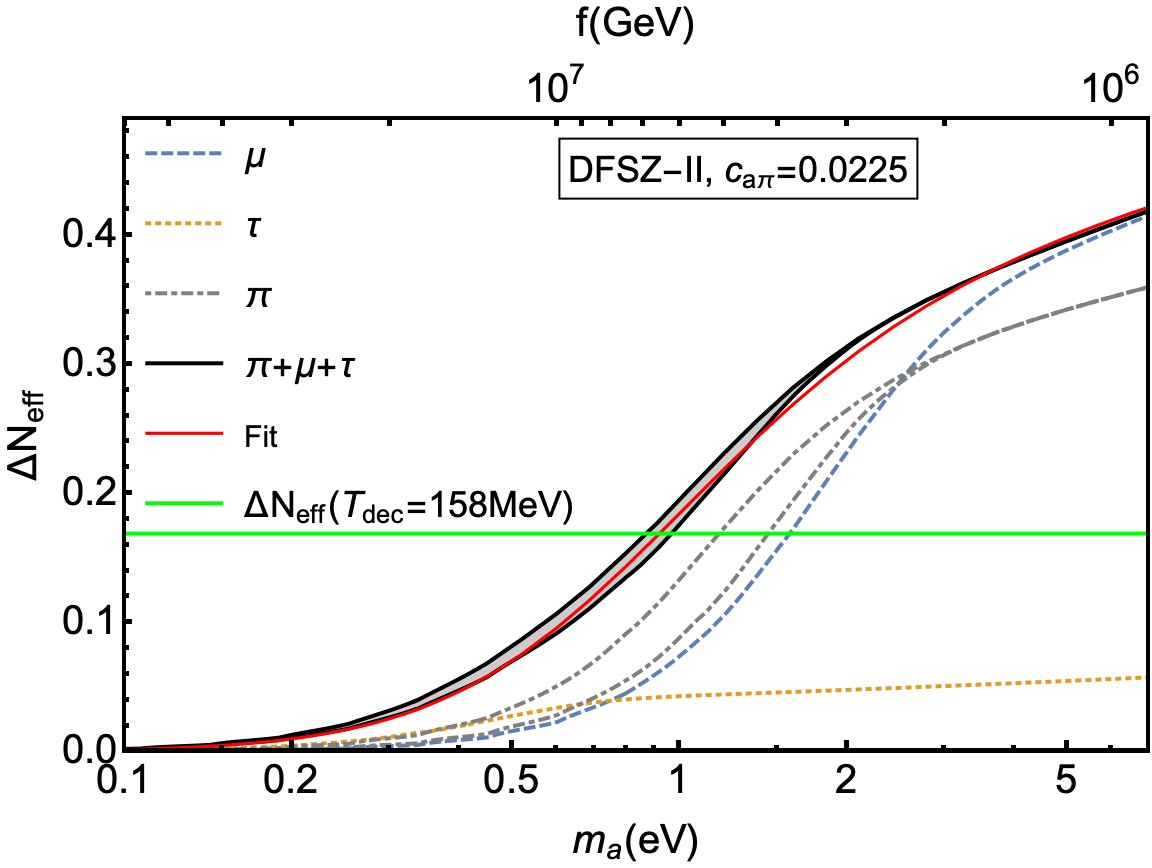}
	\caption{\small The thermal axion abundance, obtained by solving numerically the Boltzmann equation including production from pion, muon and tau scatterings, in the DFSZ-II model. We show two representative examples with $c_{a\pi}=0.225$ and $c_{a\pi}=0.0225$. We show the fitting functions (red curves) which we used in our modification to {\tt CLASS}. Solutions obtained by considering each component separately are also shown as dashed lines in the right plot (see legends). In the left plot the contributions from leptons are not visible. The green line is the value of $\nef$ if the axion decouples at $T=158$ MeV. \label{fig:boltzmannII}}
\end{figure*}

\section{Solution of the Boltzmann equation}
\label{sec:boltzmann}

Here we provide more details about the computation of $\Delta N_{\text{eff}}$ from the Boltzmann equation. 
We have solved \eqref{eq:boltzmann} including the thermally averaged pion~\eqref{eq:pionrate} and lepton scattering rates~\eqref{eq:leptonrate} and computed $\Delta N_{\text{eff}}$ by means of~\eqref{eq:deltan2}. The solution to the Boltzmann equation depends only on two parameters: $c_{a\pi}$ and $m_a$. It is also sometimes convenient to trade $c_{a\pi}$ for $\sin\beta$ (or $\tan\beta$), according to \eqref{eq:Capidef}. We include pions only below the QCD critical temperature, which we take to be  $158~\text{MeV}$ based on the latest lattice calculations~\cite{Borsanyi:2020fev}. 

We show in Figs.~\ref{fig:boltzmannI} (\ref{fig:boltzmannII}) the result of our calculation in the DFSZ-I (DFSZ-II) scenario for two representative choices $c_{a\pi}=0.225 \,(\text{left}), 0.0225\, (\text{right})$ which we have used in this work. For large values of the axion-pion coupling, the contribution from leptons is completely negligible in the mass range of interest in both scenarios, as discussed in Sec.~\ref{sec:alep}. Therefore, they share the same the same prediction for $\nef$ for large $c_{a\pi}$. Instead, for small values of $c_{a\pi}$, production from muons (and to a lesser extent from taus) is relevant only in the DFSZ-II model, as shown in Fig.~\ref{fig:boltzmannII} (right) thus yielding a larger value of $\nef$ than in the DFSZ-I model for the same values of $c_{a\pi}$ and $m_a$.

In both cases, the dependence of $\Delta N_{\text{eff}}$ on $c_{a\pi}$ is well reproduced by the function
\begin{equation}
	\label{eq:deltanfit}
	\Delta N_{\text{eff, fit}} \simeq  A\alpha^{8/3} 
	\left(1 + B \,  \alpha^{0.1/C} \right)^{C} \, .
\end{equation}
where $\alpha \equiv (m_a/\text{eV}) (c_{a\pi}/0.225)$. The behavior $ \Delta N_\text{eff} \propto \alpha^{8/3} \propto f^{-8/3}$ for small masses can be obtained analytically \cite{DEramo:2018vss}. When production from pions is dominant (DFSZ-I and DFSZ-II at large coupling), we find $A\approx 66.76, \,B\approx 228.96$ and $\,C=-0.95$ to give a good fit, shown as a red curve in Fig.~\ref{fig:boltzmannII} (left). When leptons and pions are both relevant, the values of $A, B, C$depend on the precise value of $c_{a\pi}$. For $c_{a\pi}=0.0225$, we find $A\approx 0.83, B\approx 1.7$ and $C\approx -1.34$ to provide a good fit, shown as a red curve in Fig.~\ref{fig:boltzmannII} (right). We used these fits to implement the thermal axion abundance in {\tt CLASS}.

The shaded regions in Figs.~\ref{fig:boltzmannI} and \ref{fig:boltzmannII} correspond to varying the temperature below which we include pion scatterings in the Boltzmann equation, between $158~\text{MeV}$ (lower curve) and $200~\text{MeV}$ (upper curve).
Finally, the green line corresponds to the value of $\nef$ if the axion decouples at $T=158$ MeV. Below the line, the axion does not thermalize below the QCD PT and so there could be additional contributions from pion scatterings at higher temperatures which need to be computed using different methods. In this regime we also neglect contributions from axion couplings to heavier quarks which could give a signal up to $\nef \simeq  0.05$ when decoupling happens while in weakly coupled regimes ($T\gtrsim 1$ GeV) \cite{Ferreira:2018vjj}.

\bibliography{biblio}
\bibliographystyle{BiblioStyle}

\end{document}